\documentclass[aps,prd,twocolumn,superscriptaddress,showpacs,showkeys,
nofootinbib]{revtex4-1}

\usepackage{latexsym}
\usepackage{amsmath}
\usepackage{amssymb}
\usepackage{amsfonts}

\usepackage{color}

\usepackage{supertabular} 
\usepackage{placeins}
\usepackage{epsfig}
\usepackage{graphicx}

\usepackage[hyperindex,breaklinks]{hyperref}

\hypersetup{
  colorlinks=true,
  linkcolor=blue,
  filecolor=magenta,
  urlcolor=blue,
  citecolor=blue,
}

\usepackage{soul} 

\begin{document}


\title{Does the $J^{PC}=1^{+-}$ counterpart of the $X(3872)$ exist?}

\author{Pablo G. Ortega}
\email[]{pgortega@usal.es}
\affiliation{Departamento de Física Fundamental and Instituto Universitario de 
F\'isica 
Fundamental y Matem\'aticas (IUFFyM), Universidad de Salamanca, E-37008 
Salamanca, Spain}

\author{David R. Entem}
\email[]{entem@usal.es}
\affiliation{Grupo de F\'isica Nuclear and Instituto Universitario de F\'isica 
Fundamental y Matem\'aticas (IUFFyM), Universidad de Salamanca, E-37008 
Salamanca, Spain}

\author{Francisco Fern\'andez}
\email[]{fdz@usal.es}
\affiliation{Grupo de F\'isica Nuclear and Instituto Universitario de F\'isica 
Fundamental y Matem\'aticas (IUFFyM), Universidad de Salamanca, E-37008 
Salamanca, Spain}


\date{\today}

\begin{abstract}
We explore the possible existence of the $J^{PC}=1^{+-}$ counterpart of the 
$X(3872)$ state in a coupled-channels calculation within a constituent quark 
model, with the aim of confirming the existence of the so-called 
$\tilde{X}(3872)$ state observed by the COMPASS Collaboration.
 Two states are found in the energy region of the $\tilde X(3872)$ signal, both with almost equal mixture of $c\bar c$
$2^1P_1$ state and $D^*\bar D^{(*)}$ channels: One that can be identified as the
dressed $c\bar c$ $2^1P_1$ and a bound state below the $D\bar D^*$ threshold.
We provide predictions of strong and radiative decays 
that could help to clarify the existence of such structures.
\end{abstract}



\maketitle


Since the discovery, in 2003, by the Belle Collaboration~\cite{Choi:2003ue} of 
the exotic hadron $X(3872)$, observed as a narrow peak in the $J/\psi \pi^+ 
\pi^-$ mass spectrum from the decay $B^+\longrightarrow K^{\pm} J/\psi \pi^+ 
\pi^-$, there has been plenty of states in this charmonium mass range, either 
found experimentally or predicted theoretically, that did not fit into the 
scheme predicted by the quark model.

On the theoretical side, a useful tool used to predict new hadronic states with 
heavy quarks is the Heavy-Quark Spin Symmetry (HQSS). This symmetry is based on 
the fact that, in the limit of infinite heavy quark mass, the strong interaction 
in the system is independent of the heavy quark spin. Using this symmetry, Baru 
\emph{et al.}~\cite{Baru:2016iwj} predicted the existence of three degenerate 
spin partners of the $X(3872)$ with quantum
numbers $0^{++}$, $1^{+-}$ and $2^{++}$. 
A recent systematic study of heavy-antiheavy hadronic molecules has been
performed by Dong \emph{et al.}~\cite{Dong:2021juy}, leading to similar conclusions
and predicting more than $200$ new molecules.

The $1^{+-}$ state may coincide with the signal found by the COMPASS 
Collaboration~\cite{COMPASS:2017wql} in a search for muon production of the 
$X(3872)$ through the reaction
\begin{equation}
\mu^+N\rightarrow \mu^+ X_0\pi^{\pm}N^\prime\rightarrow \mu^+ \left(J/\psi 
\pi^+\pi^-\right)\pi^{\pm}N^\prime
\end{equation}
where $X_0$ is produced by virtual photons
\begin{equation}
\gamma^*N\rightarrow X_0\pi^{\pm}N^\prime
\end{equation}

Here $N$ denotes the target nucleon, $X_0$ is an intermediate states decaying to 
$J/\psi \pi^+\pi^-$ and $N^\prime$ the unobserved recoil system.

The resulting $J/\psi \pi^+\pi^-$ invariant mass distribution shows two peaks 
with positions and widths compatible with the $\psi(2S)$ and the $X(3872)$. 
However, the shape of the second peak disagrees with previous observations of 
the $X(3872)$. The mass spectrum of the two pions from the decay of the 
$X(3872)$ shows a preference for the $X(3872)\rightarrow J/\psi \rho^0$ decay 
mode, while the shape of the $\pi^+\pi^-$ invariant mass distribution of the 
$X_0$ appears very different and it is inconsistent with the quantum numbers 
$J^{PC}=1^{++}$. Due to these differences, COMPASS Collaboration concluded that 
the observed signal is an evidence for a new charmonium-like state, dubbed the 
$\tilde{X}
(3872)$, with quantum numbers $J^{PC}=1^{+-}$ and a significance of $4.1\sigma$.
The measured mass and width of the $\tilde{X}(3872)$
are respectively $M=3860\pm 10.4$ MeV/$c^2$ and $\Gamma<51$ MeV.

To assess the possibility of the existence of this predicted state, we 
perform a calculation similar to the one done in previous studies for the 
$X(3872)$~\cite{Ortega:2009hj,Ortega:2012rs}, but in the channel 
$J^{PC}=1^{+-}$. The same model and parametrization of the aforementioned study 
are employed: a coupled-channels calculation of two and four quark sectors 
in the framework of a widely tested constituent quark 
model~\cite{Vijande:2004he,Segovia:2008zz}.
The model details can be found in Refs.~\cite{Ortega:2009hj,Ortega:2012rs} and 
references therein but, in the following, we briefly describe its main 
aspects.

The constituent quark model we use is based on the assumption that light quarks (that is, $\{u,d,s\}$ quarks)
acquire a dynamical mass as a consequence of the chiral symmetry breaking at 
some momentum scale. The breaking of the chiral symmetry entails the appearance 
of Goldstone boson exchanges between quarks. In the heavy sector (when a $c$ or $b$ quark is involved), chiral
symmetry is explicitly broken by the heavy quark mass and this type of 
interaction does not act for $QQ$ and $Qq$ quark pairs, being $Q=\{c,b\}$ and $q=\{u,d,s\}$. However, in the
proximity of the $D^*\bar D^{(*)}$ thresholds, it provides a natural way to 
incorporate pion exchanges into the $D^*\bar D^{(*)}$ dynamics through the 
light quarks.

Beyond the scale of the chiral symmetry breaking, quark-quark dynamics is 
governed by QCD perturbative effects. They are usually taken into account 
through the one-gluon exchange interaction~\cite{DeRujula:1975qlm} obtained from 
the Lagrangian,
\begin{equation}
{\mathcal L}_{qqg} = i\sqrt{4\pi\alpha_{s}} \, \bar{\psi} \gamma_{\mu} 
G^{\mu}_a \lambda^a \psi,\label{Lqqg}
\end{equation}
where $\alpha_{s}$ is the strong coupling constant, $\lambda^a$ are the $SU(3)$ 
color matrices and $G^{\mu}_a$ is the gluon field. The strong coupling constant, 
$\alpha_{s}$, has a scale dependence which allows a consistent description of 
light, strange and heavy mesons, given by

\begin{equation}
 \alpha_s(\mu) = \frac{\alpha_0}{\ln\left(\frac{\mu^2+\mu_0^2}{\Lambda_0^2}\right)}
\end{equation}
where $\mu$ is the reduced mass of the quark pair and $\mu_0$ and $\Lambda_0$ are
model parameters, which can be found in Table~\ref{tab:CQMparam}.

\begin{table}
\caption{\label{tab:CQMparam} Quark model parameters.}
\begin{tabular}{l|cc}
\hline
\hline
 Quark masses  & $m_n$ (MeV) & 313 \\
               & $m_s$ (MeV) & 555 \\
               & $m_c$ (MeV) & 1763 \\
 \hline
 Goldstone bosons & $m_\pi$ (fm$^{-1}$) & $0.70$ \\
                  & $m_\sigma$ (fm$^{-1}$) & $3.42$ \\
                  & $\Lambda$ (fm$^{-1}$) & $4.20$ \\
                  & $g^2_{\rm ch}/(4\pi)$ & $0.54$ \\
 \hline
 Confinement  & $a_c$ (MeV) & $507.4$ \\
              & $\mu_c$ (fm$^{-1}$) & $0.576$ \\
              & $\Delta$ (MeV) & $184.432$ \\
              & $a_s$ & $0.81$ \\
\hline
 OGE  & $\alpha_0$ & $2.118$ \\
      & $\Lambda_0$ (fm$^{-1}$) & $0.113$ \\
      & $\mu_0$ (MeV) & $36.976$ \\
      & $\hat r_0$ (fm) & $ 0.181$ \\
      & $\hat r_g$ (fm) & $0.259$ \\
 \hline

\end{tabular}
\end{table}

Below the chiral symmetry scale, the simplest Lagrangian is provided by the 
Instanton Liquid Model (ILM)~\cite{Diakonov:2002fq}
\begin{equation}
{\mathcal L} = \bar{\psi}(i\, {\slash\!\!\! \partial} 
-M(q^{2})U^{\gamma_{5}})\,\psi  \,,
\end{equation}
being $U^{\gamma_5} = e^{i\lambda _{a}\phi ^{a}\gamma _{5}/f_{\pi}}$ the matrix 
of Goldstone-boson fields, $\pi^a$ denotes the nine pseudoscalar fields $\{ 
\eta_0, \vec{\pi}, K_i, \eta_8\}$ with $i=1,\ldots,4$ and $M(q^2)$ is the 
dynamical mass.

An expression of this dynamical mass can be obtained from the ILM 
theory~\cite{Diakonov:2002fq}, but its behavior can be simulated by the simple 
parametrization
$M(q^2)=m_qF(q^2)$, where $m_q$ is a parameter that corresponds to the 
constituent quark mass and

\begin{equation}
F(q^2)=\left[\frac{\Lambda^2}{\Lambda^2+q^2}\right]
\end {equation}
The cut-off $\Lambda$ fixes the scale of the chiral symmetry breaking.

The Goldstone boson matrix $U^{\gamma_5}$ can be expanded in terms of the boson 
fields

\begin{equation}
U^{\gamma _{5}} = 1 + \frac{i}{f_{\pi}} \gamma^{5} \lambda^{a} \pi^{a} - 
\frac{1}{2f_{\pi}^{2}} \pi^{a} \pi^{a} + \ldots
\end{equation}
The first term of the expansion generates the constituent quark mass, the second 
one gives rise to the pseudoscalar meson-exchange interaction among quarks and 
the main contribution of the third term is on two-pion exchanges, which is 
modeled by means of a scalar-meson exchange potential. Explicit
expressions for the $\pi$ and $\sigma$ exchange potentials
can be found in Ref.~\cite{Vijande:2004he}.

The final piece of the quark-quark interaction is the confinement potential, 
which prevents the existence of colored hadrons. This potential shows a linear 
behavior but,  at a certain distance, the link between the quarks breaks up, 
giving rise to the creation of light quark-antiquark pairs. This dynamics can be 
translated into a screened potential~\cite{Born:1989iv} as

\begin{equation}
V_{\rm CON}(\vec{r}\,)=\left[-a_{c}(1-e^{-\mu_{c}r})+\Delta \right] 
(\vec{\lambda}_{q}^{c}\cdot\vec{\lambda}_{\bar{q}}^{c}) \,,
\label{eq:conf}
\end{equation}
 where $a_{c}$, $\mu_{c}$ and $\Delta$ are model parameters (see Tab.~\ref{tab:CQMparam}). At short distances, this
potential presents a linear behavior with an effective confinement strength, 
$\sigma=-a_{c}\,\mu_{c}\,(\vec{\lambda}^{c}_{i}\cdot \vec{\lambda}^{c}_{j}),$
while it becomes constant at large distances, with a plateau at $V_{\rm CON}(\vec{r}\to\infty) = \left(\Delta -a_{c}\right)
(\vec{\lambda}_{q}^{c}\cdot\vec{\lambda}_{\bar{q}}^{c})$.

To model the $J^{PC}=1^{+-}$ charmonium sector we follow the steps done in 
Ref.~\cite{Ortega:2009hj} for the $1^{++}$ sector. The full hadronic state is, 
hence, assumed to be given by

\begin{equation} \label{ec:funonda}
 | \Psi \rangle = \sum_\alpha c_\alpha | \psi_\alpha \rangle
 + \sum_\beta \chi_\beta(P) |\phi_{M_1} \phi_{M_2} \beta \rangle
\end{equation}
where $|\psi_\alpha\rangle$ are $c\bar c$ eigenstates of the two body
Hamiltonian, 
$\phi_{M_i}$ are $c\bar n$ ($\bar c n$) eigenstates describing 
the $D$ ($\bar D$) mesons, 
$|\phi_{M_1} \phi_{M_2} \beta \rangle$ is the two meson state with $\beta$ 
quantum
numbers coupled to total $J^{PC}$ quantum numbers
and $\chi_\beta(P)$ is the relative wave 
function between the two mesons in the molecule. As we always work with 
eigenstates
of the $C$-parity operator we use the usual notation in which $D\bar D^*$ is the 
right
combination of $D\bar D^*$ and $D^* \bar D$.

The meson eigenstates $\phi_C(\vec{p}_{C})$ are calculated by means of the 
two-body
Schr\"odinger equation, using the Gaussian Expansion 
Method~\cite{Hiyama:2003cu}. This method provides enough numerical accuracy for
the solution of the Schr\"odinger equation and simplifies
the subsequent evaluation of the needed matrix elements. With the aim of 
optimizing the Gaussian ranges employing a reduced number of free parameters, we 
use Gaussian trial functions whose ranges are given by a geometrical 
progression~\cite{Hiyama:2003cu}. This choice produces a dense distribution at 
short distances, enabling a better description of the dynamics mediated by short 
range potentials.

The coupling between the two and four quark sectors requires the creation of a 
light-quark pair $n\bar n$. Similar to
the strong decay process, this coupling should be in principle driven by the 
same interquark Hamiltonian which determines the spectrum. However, Ackleh 
\emph{et al.}~\cite{Ackleh:1996yt} have shown that the quark pair creation 
$^3P_0$
model~\cite{LeYaouanc:1972vsx} gives similar results to the microscopic 
calculation. The model assumes that the pair creation Hamiltonian is
\begin{equation} 
\mathcal{H}=g \int d^3x \,\, \bar \psi(x) \psi(x)
\end{equation}
which in the non-relativistic reduction is equivalent to the transition
operator~\cite{Bonnaz:1999zj}
\begin{eqnarray}
T&=&-3\sqrt{2}\gamma'\sum_\mu \int d^3 p d^3p' \,\delta^{(3)}(p+p')\times 
\nonumber\\  
&\times&
\left[ \mathcal Y_1\left(\frac{p-p'}{2}\right) b_\mu^\dagger(p)
d_{\bar \mu}^\dagger(p') \right]^{C=1,I=0,S=1,J=0}
\label{TBon}
\end{eqnarray}
where $\mu$ ($\bar \mu$) are the quark (antiquark) quantum numbers and
$\gamma'=2^{5/2} \pi^{1/2}\gamma$ with $\gamma= \frac{g}{2m}$ is a dimensionless 
constant 
that gives the strength of 
the $n\bar n$ pair creation from the vacuum.
From this operator we define the transition
potential $V_{\beta \alpha}(P)$ within the $^3 P_0$ model 
as~\cite{Kalashnikova:2005ui} 
\begin{equation}
\label{Vab}
	\langle \phi_{M_1} \phi_{M_2} \beta | T | \psi_\alpha \rangle =
	P \, V_{\beta \alpha}(P) \,\delta^{(3)}(\vec P_{\mbox{cm}})
\end{equation}
where $P$ is the relative three-momentum of the two meson state.

\begin{figure}[htb!]
 \includegraphics[width=.5\textwidth]{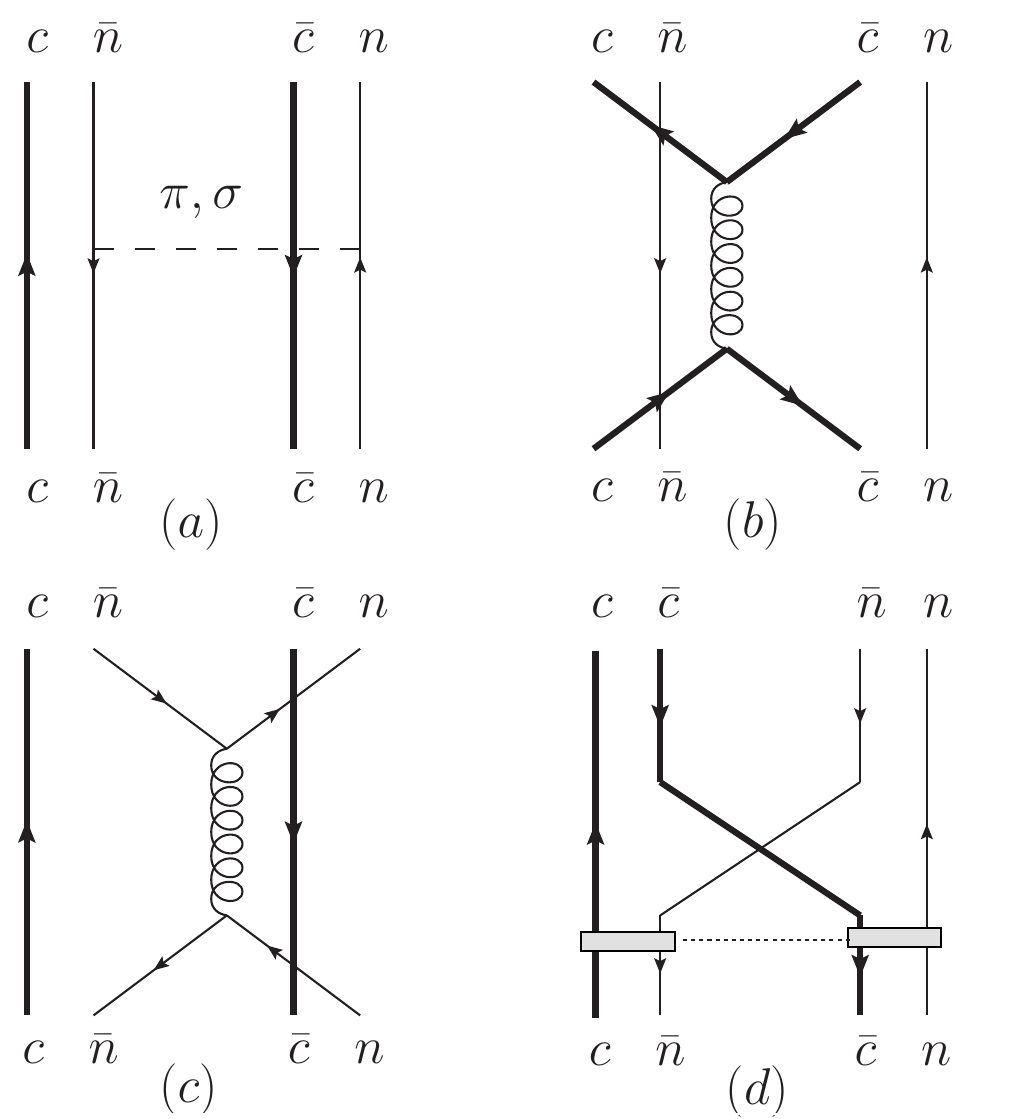}
 \caption{\label{fig1} Diagrams for the quark-quark interactions considered in the RGM Hamiltonian $H_{\beta'\beta}$ of Eq.~\eqref{fig:coupleE}: a) Direct exchange of $\pi$ and $\sigma$, b) and c) annihilation diagrams through a gluon and d) quark rearrangement diagrams, where the gray band represents the sum of interactions between quarks of different clusters  and the dotted line represents contributions from $\pi$, $\sigma$, confinement and one-gluon-exchange potentials. }
\end{figure}

Using the wave function from Eq.~\eqref{ec:funonda} and the coupling
Eq.~\eqref{Vab}, we arrive to the coupled equations
\begin{align}
\label{fig:coupleE}
M_\alpha \,c_\alpha + \sum_\beta\int V_{\alpha \beta}(P) \chi_{\beta} (P)\,P^2
\,dP =& E \,c_\alpha 
\nonumber \\
\sum_{\beta} \int H_{\beta'\beta} (P',P)
\chi_\beta(P) \, P^2 \, dP +  
& \\
+\sum_\alpha V_{\beta' \alpha}(P') c_\alpha =&E \,\chi_{\beta'}(P')\nonumber
\end{align}
where $M_\alpha$ are the masses of the bare $c\bar c$ mesons and
$H_{\beta'\beta}$ is the RGM Hamiltonian for the two meson
states obtained from the $q\bar q$ interaction~\cite{Wheeler:1937zza,Ortega:2021zgk}.
This includes direct diagrams that have contributions from
one-pion exchange (OPE), one-sigma exchange (OSE) and the anihilation through a gluon, and
also rearrangement diagrams~\cite{Ortega:2009hj} that have contributions from OPE, OSE, confinement
and one-gluon exchange (see Fig.~\ref{fig1}).

Solving these coupled equations, we can describe both the renormalization of the 
bare $c\bar c$ states due to the presence of nearby
meson-meson thresholds and the generation of new states 
through the underlying $qq$ interaction that generates the residual
meson-meson interaction and the additional interaction from the coupling with intermediate $c\bar c$, as it is the case 
for the $X(3872)$ in our model~\cite{Ortega:2009hj}.

The present calculation of the $J^{PC}=1^{+-}$ sector includes the $2^1P_1$ 
$c\bar c$ state corresponding to the $h_c(2P)$ meson with bare 
mass of $3955.7$ MeV/$c^2$, coupled to the $I=0$ $D\bar D^*$ and $D^*\bar D^*$ 
molecular states in $^3S_1$ and $^3D_1$ partial waves.
 The $D^*\bar D^*$ is negligible for the $J^{PC}=1^{++}$ sector as it is only allowed in a relative $^5D_1$ partial wave but, as we
will see, this channel has a sizable effect for the  $J^{PC}=1^{+-}$ sector.
All the parameters of the model are constrained from previous analysis of the 
heavy meson phenomenology, and the value of the $^3P_0$ $\gamma$ parameter is 
taken from previous studies of the charmonium $3.9$ GeV energy 
region~\cite{Ortega:2009hj,Ortega:2017qmg,Ortega:2012rs}, thus, in that sense, 
it is a parameter-free calculation. However, being a phenomenological model,
predictions have to be taken with care, since systematic uncertainties cannot be
evaluated, which could be significant for this specific problem. Here, we hope this is not
the case and just give the predictions of our model as a possible indication of
the existence of the $\tilde X$ state.

We find two states which, provisionally, we will call \emph{state A} and 
\emph{state B}. The mass and width of both states are shown in 
Table~\ref{tab:t1}.

\begin{table}
\caption{\label{tab:t1} Masses (in MeV/$c^2$) and decay widths (in MeV) of the states A and B. }
\begin{tabular}{ccc|cccccc}
\hline
\hline
 State  & Mass & Width & $D\bar D^*$ & $\omega\eta_c$ & $\eta J/\psi$ &$J/\psi \pi \pi$ & 
$\gamma\eta_c$ & $\gamma\eta_c^\prime$ \\
 \hline
 $A$ & $3868$& $1.35$ & $0.0$ & $0.64$ & $0.56$ & $0.017$ & $0.069$ & $0.062$  \\
 $B$  & $3877$& $45.07$ & $43.2$ & $0.90$ & $0.81$ & $0.026$ & $0.064$ & $0.062$  \\
\hline
\end{tabular}
\end{table}

\begin{figure}[t]
 \includegraphics[width=0.45\textwidth]{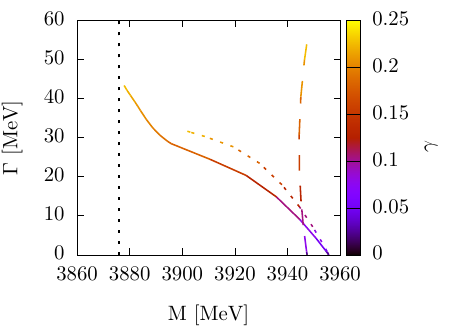}
 \caption{\label{fig:gamma} Evolution of the  dressed $2^1P_1$ $c\bar c$  coupled to $D\bar D^*+D^*\bar D^*$ channels (solid), the dressed $2^1P_1$ $c\bar c$ coupled solely to $D\bar D^*$ channel (dotted)  and
$2^3P_1$ $c\bar c$ (long-dashed) with increasing values of the $^3P_0$ strength 
parameter $\gamma$. The dotted vertical line represents the $D\bar D^*$ 
threshold. }
\end{figure}

The components of both states are shown in Table~\ref{tab:t2}.
The first striking result is that both states share almost the same proportion 
of $h_c(2P)$ and $D^*\bar D^{(*)}$ components which makes its interpretation
challenging. In the calculation of the $J^{PC}=1^{++}$ sector done in 
Ref.~\cite{Ortega:2012rs} we obtained two states: the first one with $87\%$ of 
$D^0\bar D^{*\,0}$ component and a second state with more than $60\%$ of the
 $2^3P_1$ $c\bar c$ state. Thus, in that case, the coupling of $c\bar c$ and 
$D\bar D^*$ channels produced a clear extra state, the $X(3872)$, and a 
renormalized $c\bar c$ state, assigned to the $X(3940)$ resonance, whose 
dressing evolution with increasing $\gamma$ is shown in the Fig.~\ref{fig:gamma}.

The case of the $J^{PC}=1^{+-}$ is different. Besides an extra bound state, called state $A$,
the $2^1P_1$ $c\bar c$ moves towards 
the $D\bar D^*$ threshold. The $D^*\bar D^*$ channel, although it is $280$ MeV above, produces a sizable attraction effect, which is not present in the $J^{PC}=1^{++}$ sector as there only the relative $^5D_1$ partial wave is opened.
In Fig.~\ref{fig:gamma}, we show the evolution of the dressed $2^1P_1$ $c\bar c$
state with increasing $\gamma$ values to make clearer the dressing mechanism in 
comparison with the $J^{PC}=1^{++}$. With the aim of showing the effect of the $D^*\bar D^*$ channel in the dynamics of the $1^{+-}$ sector, we also add the evolution of the dressed $2^1P_1$ $c\bar c$ state
coupled solely to the $D\bar D^*$.

\begin{table}
\caption{\label{tab:t2} Probabilities (in $\%~$) of the three coupled channels 
for the two different states described in the text.}
\begin{tabular}{cccc}
\hline
\hline
 State  & $h_c(2P)$ & $D\bar D^*$ & $D^*\bar D^*$ \\
 \hline
 $A$  & $49.7$ & $45.5$ & $4.8$ \\
 $B$  & $44.7$ & $50.0$ & $5.3$ \\
\hline
\end{tabular}
\end{table}

Our results show that the coupling with the bare $c\bar c$ states modifies the
consequences of HQSS~\cite{doi:10.1063/1.4949442}. This symmetry predicts that the $D\bar D^*$ 
and the $D^*\bar D^*$ channels includes a mix of spin-singlet and spin-triplet 
states with equal weights. However, the 
$h_c(2^1P_1)$ $c\bar c$ state only couples, via the $^3P_0$ mechanism, with the 
spin-singlet part of those channels and, therefore, it enhances this component (see 
Table~\ref{tab:t3}). This spin structure is important to understand the decay 
channels of both states.

\begin{table}
\caption{\label{tab:t3} Molecular components with
$c\bar c$ coupled to spin singlet and triplet for the two 
different states described in the text.}
\begin{tabular}{ccc}
\hline
\hline
 State  & Singlet($\%$) & Triplet ($\%$)  \\
 \hline
 $A$  & $79.2$ & $20.8$  \\
 $B$  & $75.6$ & $24.4$ \\
\hline
\end{tabular}
\end{table}

In Table~\ref{tab:t1} we also show the widths of both states $A$ and $B$ 
corresponding to the decay channels discussed below.

The state $B$ is above the $D\bar D^*$ threshold and, therefore, its main 
decay channel is $D\bar D^*$. The state $A$ is below the $D\bar D^*$ threshold. Then, its decays to $D\bar 
D^*$ or $D^*\bar D^*$ channels are forbidden. 

On top of that, some relevant decay channels of both states are the $\omega 
\eta_c$ and the $\eta J/\psi$. These decays go through exchange diagrams where 
the quarks are rearranged inside the $D^* \bar D^{(*)}$ channels (see Ref.~\cite{Ortega:2009hj} for details). Even if
exchanges diagrams are usually small compared to the direct diagrams in coupled-channels 
calculations, the decay width may be sufficiently large to be measured.
The $\eta J/\psi$ decay goes through the spin-triplet component, whereas the 
$\omega \eta_c$ through the spin-singlet one. However, the widths corresponding 
to both decays are of the same order of magnitude, because the phase space of the $\eta J/\psi$ 
channel is larger than the $\omega \eta_c$ one.

The $\tilde X(3872)$ particle was spotted in the $J/\psi\pi^+\pi^-$ invariant mass spectrum, so 
it is worth discussing the details of this decay channel. 
Thinking naively, in view of the dominance of the spin-singlet components, one could 
expect that the states found must decay through their $h_c(2^1P_1)$ component. However, the 
$h_c(2^1P_1)\to J/\psi \pi^+ \pi^-$ decay is a spin-flip hadronic transition 
and, therefore, its probability should be low. Alternatively, the existence of the 
spin-triplet component may enhance the $J/\psi\pi^+\pi^-$ branching through the molecular transition
$D^+\bar D^{(*)}\to J/\psi f_0$ in P wave, $f_0$ decaying into $\pi^+ \pi^-$ where $f_0$ is a $0^{++}$ 
meson with a width of $455$ MeV.

Regarding radiative decays, in this work we consider them given by the $c\bar c$ components
of the states, and so we don't consider the decay throught the molecular component. Since the $h_c(2^1P_1)$ has negative C parity, it is
very likely that it decays into a photon plus a pseudoscalar meson such $\eta_c$ 
and $\eta_c^\prime$. Looking at the PDG data~\cite{ParticleDataGroup:2020ssz} of
the $h_c(1^1P_1)$ state, for comparison, the widths of these decays may be 
competitive with the hadronic one.

As seen in Table~\ref{tab:t1}, the widths of the $J/\psi \pi^+ \pi^-$ hadronic decay and the 
radiative decays $\gamma\eta_c$ and $\gamma\eta_c^\prime$ are of the same order 
of magnitude, at keV range. However, in the case of the state $A$, 
it represents  $\sim\!\!1\,\%$
of the total width, whereas the contribution to the total width is negligible 
for the state $B$.

Of course, the way to distinguish between the two states,
is that state $B$ can decay into $D\bar D^*$ which is
responsible of almost all the decay width, and should be more accessible in this 
channel. However, if the production of the $\tilde X$ is not significative, its signal may be hidden in the $X(3872)$ threshold
enhancement in the $D\bar D^*$ channel, so the radiative or strong decays 
mentioned earlier could be more promising. Taken into account our predictions at 
Table~\ref{tab:t1}, we encourage experimentalists to search for such 
resonances in the $\omega\eta_c$ and $\eta J/\psi$ channels.

As summary,  we have performed a coupled channels calculation within the 
constituent quark model of the charmonium $J^{PC}=1^{+-}$ sector in order to 
confirm the possible existence of the partner of the $X(3872)$ named 
$\tilde{X}(3872)$. 

We found in this energy region two almost degenerate states with masses 
$M_A= 3868$ MeV/$c^2$ and $M_B=3877$ MeV/$c^2$. Considering strong and radiative 
decays, the width of the $A$ state is $1.35$ MeV whereas the $B$ state has a 
width of $45.07$ MeV.

Our results are in line with previous studies, which usually employ HQSS to predict partners of the $X(3872)$. We have already cited Ref.~\cite{Dong:2021juy} where the authors predict a $1^{+-}$ bound state with a mass between $3874.4$ and $3839.8$ MeV. Similarly, Ref.~\cite{Yan:2021tcp} finds a negative C-parity virtual state at $3860.0\pm 10.4$ MeV,  Ref.~\cite{Gamermann:2007fi} predicts a $1^{+-}$ state at a mass of $3840.69$ MeV, while Ref.~\cite{Wang:2020dgr}, using QCD sum rules, finds a $1^{+-}$ state at $3.89\pm0.09$ GeV together with the $1^{++}$ state.
The difference with our calculation is that we include the coupling with $c\bar c$ bare states which, together with the molecular state, leads to the emergence of a second pole due to the dressing of the $2^1P_1$ $c\bar c$ state in the same energy region.

Both state are mainly a composite of $h_c(2^1P_1)$ $c\bar c$ states and 
$D^*\bar D^{(*)}$ molecule with almost the same probability.
With the scarce data available, it is difficult to experimentally identify these 
states as a renormalized  $h_c(2^1P_1)$ and an extra state $\tilde{X}(3872)$. We 
suggest possible ways to obtain more precise measurements of their properties 
via radiative and strong decays.


\begin{acknowledgments}
The authors would like to thank Prof.~F.-K.~Guo for fruitful discussions and useful comments.

This work has been partially funded by 
Ministerio de Ciencia, Innovaci\'on y Universidades
under Contract No. PID2019-105439GB-C22/AEI/10.13039/501100011033,
and by the EU Horizon 2020 research and innovation program, STRONG-2020 project, 
under grant agreement No. 824093.
\end{acknowledgments}


\bibliography{Xtilde}

\begin{thebibliography}{24}%
\makeatletter
\providecommand \@ifxundefined [1]{%
 \@ifx{#1\undefined}
}%
\providecommand \@ifnum [1]{%
 \ifnum #1\expandafter \@firstoftwo
 \else \expandafter \@secondoftwo
 \fi
}%
\providecommand \@ifx [1]{%
 \ifx #1\expandafter \@firstoftwo
 \else \expandafter \@secondoftwo
 \fi
}%
\providecommand \natexlab [1]{#1}%
\providecommand \enquote  [1]{``#1''}%
\providecommand \bibnamefont  [1]{#1}%
\providecommand \bibfnamefont [1]{#1}%
\providecommand \citenamefont [1]{#1}%
\providecommand \href@noop [0]{\@secondoftwo}%
\providecommand \href [0]{\begingroup \@sanitize@url \@href}%
\providecommand \@href[1]{\@@startlink{#1}\@@href}%
\providecommand \@@href[1]{\endgroup#1\@@endlink}%
\providecommand \@sanitize@url [0]{\catcode `\\12\catcode `\$12\catcode
  `\&12\catcode `\#12\catcode `\^12\catcode `\_12\catcode `\%12\relax}%
\providecommand \@@startlink[1]{}%
\providecommand \@@endlink[0]{}%
\providecommand \url  [0]{\begingroup\@sanitize@url \@url }%
\providecommand \@url [1]{\endgroup\@href {#1}{\urlprefix }}%
\providecommand \urlprefix  [0]{URL }%
\providecommand \Eprint [0]{\href }%
\providecommand \doibase [0]{http://dx.doi.org/}%
\providecommand \selectlanguage [0]{\@gobble}%
\providecommand \bibinfo  [0]{\@secondoftwo}%
\providecommand \bibfield  [0]{\@secondoftwo}%
\providecommand \translation [1]{[#1]}%
\providecommand \BibitemOpen [0]{}%
\providecommand \bibitemStop [0]{}%
\providecommand \bibitemNoStop [0]{.\EOS\space}%
\providecommand \EOS [0]{\spacefactor3000\relax}%
\providecommand \BibitemShut  [1]{\csname bibitem#1\endcsname}%
\let\auto@bib@innerbib\@empty
\bibitem [{\citenamefont {Choi}\ \emph {et~al.}(2003)\citenamefont {Choi} \emph
  {et~al.}}]{Choi:2003ue}%
  \BibitemOpen
  \bibfield  {author} {\bibinfo {author} {\bibfnamefont {S.~K.}\ \bibnamefont
  {Choi}} \emph {et~al.} (\bibinfo {collaboration} {Belle}),\ }\href {\doibase
  10.1103/PhysRevLett.91.262001} {\bibfield  {journal} {\bibinfo  {journal}
  {Phys. Rev. Lett.}\ }\textbf {\bibinfo {volume} {91}},\ \bibinfo {pages}
  {262001} (\bibinfo {year} {2003})},\ \Eprint
  {http://arxiv.org/abs/hep-ex/0309032} {arXiv:hep-ex/0309032 [hep-ex]}
  \BibitemShut {NoStop}%
\bibitem [{\citenamefont {Baru}\ \emph {et~al.}(2016)\citenamefont {Baru},
  \citenamefont {Epelbaum}, \citenamefont {Filin}, \citenamefont {Hanhart},
  \citenamefont {Mei\ss{}ner},\ and\ \citenamefont {Nefediev}}]{Baru:2016iwj}%
  \BibitemOpen
  \bibfield  {author} {\bibinfo {author} {\bibfnamefont {V.}~\bibnamefont
  {Baru}}, \bibinfo {author} {\bibfnamefont {E.}~\bibnamefont {Epelbaum}},
  \bibinfo {author} {\bibfnamefont {A.~A.}\ \bibnamefont {Filin}}, \bibinfo
  {author} {\bibfnamefont {C.}~\bibnamefont {Hanhart}}, \bibinfo {author}
  {\bibfnamefont {U.-G.}\ \bibnamefont {Mei\ss{}ner}}, \ and\ \bibinfo {author}
  {\bibfnamefont {A.~V.}\ \bibnamefont {Nefediev}},\ }\href {\doibase
  10.1016/j.physletb.2016.10.008} {\bibfield  {journal} {\bibinfo  {journal}
  {Phys. Lett. B}\ }\textbf {\bibinfo {volume} {763}},\ \bibinfo {pages} {20}
  (\bibinfo {year} {2016})},\ \Eprint {http://arxiv.org/abs/1605.09649}
  {arXiv:1605.09649 [hep-ph]} \BibitemShut {NoStop}%
\bibitem [{\citenamefont {Dong}\ \emph {et~al.}(2021)\citenamefont {Dong},
  \citenamefont {Guo},\ and\ \citenamefont {Zou}}]{Dong:2021juy}%
  \BibitemOpen
  \bibfield  {author} {\bibinfo {author} {\bibfnamefont {X.-K.}\ \bibnamefont
  {Dong}}, \bibinfo {author} {\bibfnamefont {F.-K.}\ \bibnamefont {Guo}}, \
  and\ \bibinfo {author} {\bibfnamefont {B.-S.}\ \bibnamefont {Zou}},\ }\href
  {\doibase 10.13725/j.cnki.pip.2021.02.001} {\bibfield  {journal} {\bibinfo
  {journal} {Progr. Phys.}\ }\textbf {\bibinfo {volume} {41}},\ \bibinfo
  {pages} {65} (\bibinfo {year} {2021})},\ \Eprint
  {http://arxiv.org/abs/2101.01021} {arXiv:2101.01021 [hep-ph]} \BibitemShut
  {NoStop}%
\bibitem [{\citenamefont {Aghasyan}\ \emph {et~al.}(2018)\citenamefont
  {Aghasyan} \emph {et~al.}}]{COMPASS:2017wql}%
  \BibitemOpen
  \bibfield  {author} {\bibinfo {author} {\bibfnamefont {M.}~\bibnamefont
  {Aghasyan}} \emph {et~al.} (\bibinfo {collaboration} {COMPASS}),\ }\href
  {\doibase 10.1016/j.physletb.2018.07.008} {\bibfield  {journal} {\bibinfo
  {journal} {Phys. Lett. B}\ }\textbf {\bibinfo {volume} {783}},\ \bibinfo
  {pages} {334} (\bibinfo {year} {2018})},\ \Eprint
  {http://arxiv.org/abs/1707.01796} {arXiv:1707.01796 [hep-ex]} \BibitemShut
  {NoStop}%
\bibitem [{\citenamefont {Ortega}\ \emph {et~al.}(2010)\citenamefont {Ortega},
  \citenamefont {Segovia}, \citenamefont {Entem},\ and\ \citenamefont
  {Fernandez}}]{Ortega:2009hj}%
  \BibitemOpen
  \bibfield  {author} {\bibinfo {author} {\bibfnamefont {P.~G.}\ \bibnamefont
  {Ortega}}, \bibinfo {author} {\bibfnamefont {J.}~\bibnamefont {Segovia}},
  \bibinfo {author} {\bibfnamefont {D.~R.}\ \bibnamefont {Entem}}, \ and\
  \bibinfo {author} {\bibfnamefont {F.}~\bibnamefont {Fernandez}},\ }\href
  {\doibase 10.1103/PhysRevD.81.054023} {\bibfield  {journal} {\bibinfo
  {journal} {Phys. Rev. D}\ }\textbf {\bibinfo {volume} {81}},\ \bibinfo
  {pages} {054023} (\bibinfo {year} {2010})},\ \Eprint
  {http://arxiv.org/abs/0907.3997} {arXiv:0907.3997 [hep-ph]} \BibitemShut
  {NoStop}%
\bibitem [{\citenamefont {Ortega}\ \emph {et~al.}(2013)\citenamefont {Ortega},
  \citenamefont {Entem},\ and\ \citenamefont {Fernandez}}]{Ortega:2012rs}%
  \BibitemOpen
  \bibfield  {author} {\bibinfo {author} {\bibfnamefont {P.~G.}\ \bibnamefont
  {Ortega}}, \bibinfo {author} {\bibfnamefont {D.~R.}\ \bibnamefont {Entem}}, \
  and\ \bibinfo {author} {\bibfnamefont {F.}~\bibnamefont {Fernandez}},\ }\href
  {\doibase 10.1088/0954-3899/40/6/065107} {\bibfield  {journal} {\bibinfo
  {journal} {J. Phys. G}\ }\textbf {\bibinfo {volume} {40}},\ \bibinfo {pages}
  {065107} (\bibinfo {year} {2013})},\ \Eprint {http://arxiv.org/abs/1205.1699}
  {arXiv:1205.1699 [hep-ph]} \BibitemShut {NoStop}%
\bibitem [{\citenamefont {Vijande}\ \emph {et~al.}(2005)\citenamefont
  {Vijande}, \citenamefont {Fernandez},\ and\ \citenamefont
  {Valcarce}}]{Vijande:2004he}%
  \BibitemOpen
  \bibfield  {author} {\bibinfo {author} {\bibfnamefont {J.}~\bibnamefont
  {Vijande}}, \bibinfo {author} {\bibfnamefont {F.}~\bibnamefont {Fernandez}},
  \ and\ \bibinfo {author} {\bibfnamefont {A.}~\bibnamefont {Valcarce}},\
  }\href {\doibase 10.1088/0954-3899/31/5/017} {\bibfield  {journal} {\bibinfo
  {journal} {J. Phys.}\ }\textbf {\bibinfo {volume} {G31}},\ \bibinfo {pages}
  {481} (\bibinfo {year} {2005})},\ \Eprint
  {http://arxiv.org/abs/hep-ph/0411299} {arXiv:hep-ph/0411299 [hep-ph]}
  \BibitemShut {NoStop}%
\bibitem [{\citenamefont {Segovia}\ \emph {et~al.}(2008)\citenamefont
  {Segovia}, \citenamefont {Yasser}, \citenamefont {Entem},\ and\ \citenamefont
  {Fernandez}}]{Segovia:2008zz}%
  \BibitemOpen
  \bibfield  {author} {\bibinfo {author} {\bibfnamefont {J.}~\bibnamefont
  {Segovia}}, \bibinfo {author} {\bibfnamefont {A.~M.}\ \bibnamefont {Yasser}},
  \bibinfo {author} {\bibfnamefont {D.~R.}\ \bibnamefont {Entem}}, \ and\
  \bibinfo {author} {\bibfnamefont {F.}~\bibnamefont {Fernandez}},\ }\href
  {\doibase 10.1103/PhysRevD.78.114033} {\bibfield  {journal} {\bibinfo
  {journal} {Phys. Rev.}\ }\textbf {\bibinfo {volume} {D78}},\ \bibinfo {pages}
  {114033} (\bibinfo {year} {2008})}\BibitemShut {NoStop}%
\bibitem [{\citenamefont {De~Rujula}\ \emph {et~al.}(1975)\citenamefont
  {De~Rujula}, \citenamefont {Georgi},\ and\ \citenamefont
  {Glashow}}]{DeRujula:1975qlm}%
  \BibitemOpen
  \bibfield  {author} {\bibinfo {author} {\bibfnamefont {A.}~\bibnamefont
  {De~Rujula}}, \bibinfo {author} {\bibfnamefont {H.}~\bibnamefont {Georgi}}, \
  and\ \bibinfo {author} {\bibfnamefont {S.~L.}\ \bibnamefont {Glashow}},\
  }\href {\doibase 10.1103/PhysRevD.12.147} {\bibfield  {journal} {\bibinfo
  {journal} {Phys. Rev. D}\ }\textbf {\bibinfo {volume} {12}},\ \bibinfo
  {pages} {147} (\bibinfo {year} {1975})}\BibitemShut {NoStop}%
\bibitem [{\citenamefont {Diakonov}(2003)}]{Diakonov:2002fq}%
  \BibitemOpen
  \bibfield  {author} {\bibinfo {author} {\bibfnamefont {D.}~\bibnamefont
  {Diakonov}},\ }\href {\doibase 10.1016/S0146-6410(03)90014-7} {\bibfield
  {journal} {\bibinfo  {journal} {Prog. Part. Nucl. Phys.}\ }\textbf {\bibinfo
  {volume} {51}},\ \bibinfo {pages} {173} (\bibinfo {year} {2003})},\ \Eprint
  {http://arxiv.org/abs/hep-ph/0212026} {arXiv:hep-ph/0212026 [hep-ph]}
  \BibitemShut {NoStop}%
\bibitem [{\citenamefont {Born}\ \emph {et~al.}(1989)\citenamefont {Born},
  \citenamefont {Laermann}, \citenamefont {Pirch}, \citenamefont {Walsh},\ and\
  \citenamefont {Zerwas}}]{Born:1989iv}%
  \BibitemOpen
  \bibfield  {author} {\bibinfo {author} {\bibfnamefont {K.~D.}\ \bibnamefont
  {Born}}, \bibinfo {author} {\bibfnamefont {E.}~\bibnamefont {Laermann}},
  \bibinfo {author} {\bibfnamefont {N.}~\bibnamefont {Pirch}}, \bibinfo
  {author} {\bibfnamefont {T.~F.}\ \bibnamefont {Walsh}}, \ and\ \bibinfo
  {author} {\bibfnamefont {P.~M.}\ \bibnamefont {Zerwas}},\ }\href {\doibase
  10.1103/PhysRevD.40.1653} {\bibfield  {journal} {\bibinfo  {journal} {Phys.
  Rev. D}\ }\textbf {\bibinfo {volume} {40}},\ \bibinfo {pages} {1653}
  (\bibinfo {year} {1989})}\BibitemShut {NoStop}%
\bibitem [{\citenamefont {Hiyama}\ \emph {et~al.}(2003)\citenamefont {Hiyama},
  \citenamefont {Kino},\ and\ \citenamefont {Kamimura}}]{Hiyama:2003cu}%
  \BibitemOpen
  \bibfield  {author} {\bibinfo {author} {\bibfnamefont {E.}~\bibnamefont
  {Hiyama}}, \bibinfo {author} {\bibfnamefont {Y.}~\bibnamefont {Kino}}, \ and\
  \bibinfo {author} {\bibfnamefont {M.}~\bibnamefont {Kamimura}},\ }\href
  {\doibase 10.1016/S0146-6410(03)90015-9} {\bibfield  {journal} {\bibinfo
  {journal} {Prog. Part. Nucl. Phys.}\ }\textbf {\bibinfo {volume} {51}},\
  \bibinfo {pages} {223} (\bibinfo {year} {2003})}\BibitemShut {NoStop}%
\bibitem [{\citenamefont {Ackleh}\ \emph {et~al.}(1996)\citenamefont {Ackleh},
  \citenamefont {Barnes},\ and\ \citenamefont {Swanson}}]{Ackleh:1996yt}%
  \BibitemOpen
  \bibfield  {author} {\bibinfo {author} {\bibfnamefont {E.~S.}\ \bibnamefont
  {Ackleh}}, \bibinfo {author} {\bibfnamefont {T.}~\bibnamefont {Barnes}}, \
  and\ \bibinfo {author} {\bibfnamefont {E.~S.}\ \bibnamefont {Swanson}},\
  }\href {\doibase 10.1103/PhysRevD.54.6811} {\bibfield  {journal} {\bibinfo
  {journal} {Phys. Rev. D}\ }\textbf {\bibinfo {volume} {54}},\ \bibinfo
  {pages} {6811} (\bibinfo {year} {1996})},\ \Eprint
  {http://arxiv.org/abs/hep-ph/9604355} {arXiv:hep-ph/9604355} \BibitemShut
  {NoStop}%
\bibitem [{\citenamefont {Le~Yaouanc}\ \emph {et~al.}(1973)\citenamefont
  {Le~Yaouanc}, \citenamefont {Oliver}, \citenamefont {Pene},\ and\
  \citenamefont {Raynal}}]{LeYaouanc:1972vsx}%
  \BibitemOpen
  \bibfield  {author} {\bibinfo {author} {\bibfnamefont {A.}~\bibnamefont
  {Le~Yaouanc}}, \bibinfo {author} {\bibfnamefont {L.}~\bibnamefont {Oliver}},
  \bibinfo {author} {\bibfnamefont {O.}~\bibnamefont {Pene}}, \ and\ \bibinfo
  {author} {\bibfnamefont {J.~C.}\ \bibnamefont {Raynal}},\ }\href {\doibase
  10.1103/PhysRevD.8.2223} {\bibfield  {journal} {\bibinfo  {journal} {Phys.
  Rev. D}\ }\textbf {\bibinfo {volume} {8}},\ \bibinfo {pages} {2223} (\bibinfo
  {year} {1973})}\BibitemShut {NoStop}%
\bibitem [{\citenamefont {Bonnaz}\ and\ \citenamefont
  {Silvestre-Brac}(1999)}]{Bonnaz:1999zj}%
  \BibitemOpen
  \bibfield  {author} {\bibinfo {author} {\bibfnamefont {R.}~\bibnamefont
  {Bonnaz}}\ and\ \bibinfo {author} {\bibfnamefont {B.}~\bibnamefont
  {Silvestre-Brac}},\ }\href {\doibase 10.1007/s006010050128} {\bibfield
  {journal} {\bibinfo  {journal} {Few Body Syst.}\ }\textbf {\bibinfo {volume}
  {27}},\ \bibinfo {pages} {163} (\bibinfo {year} {1999})}\BibitemShut
  {NoStop}%
\bibitem [{\citenamefont {Kalashnikova}(2005)}]{Kalashnikova:2005ui}%
  \BibitemOpen
  \bibfield  {author} {\bibinfo {author} {\bibfnamefont {Y.~S.}\ \bibnamefont
  {Kalashnikova}},\ }\href {\doibase 10.1103/PhysRevD.72.034010} {\bibfield
  {journal} {\bibinfo  {journal} {Phys. Rev. D}\ }\textbf {\bibinfo {volume}
  {72}},\ \bibinfo {pages} {034010} (\bibinfo {year} {2005})},\ \Eprint
  {http://arxiv.org/abs/hep-ph/0506270} {arXiv:hep-ph/0506270} \BibitemShut
  {NoStop}%
\bibitem [{\citenamefont {Wheeler}(1937)}]{Wheeler:1937zza}%
  \BibitemOpen
  \bibfield  {author} {\bibinfo {author} {\bibfnamefont {J.~A.}\ \bibnamefont
  {Wheeler}},\ }\href {\doibase 10.1103/PhysRev.52.1083} {\bibfield  {journal}
  {\bibinfo  {journal} {Phys. Rev.}\ }\textbf {\bibinfo {volume} {52}},\
  \bibinfo {pages} {1083} (\bibinfo {year} {1937})}\BibitemShut {NoStop}%
\bibitem [{\citenamefont {Ortega}\ \emph {et~al.}(2021)\citenamefont {Ortega},
  \citenamefont {Entem},\ and\ \citenamefont {Fernandez}}]{Ortega:2021zgk}%
  \BibitemOpen
  \bibfield  {author} {\bibinfo {author} {\bibfnamefont {P.~G.}\ \bibnamefont
  {Ortega}}, \bibinfo {author} {\bibfnamefont {D.~R.}\ \bibnamefont {Entem}}, \
  and\ \bibinfo {author} {\bibfnamefont {F.}~\bibnamefont {Fernandez}},\ }\href
  {\doibase 10.3390/sym13091600} {\bibfield  {journal} {\bibinfo  {journal}
  {Symmetry}\ }\textbf {\bibinfo {volume} {13}},\ \bibinfo {pages} {1600}
  (\bibinfo {year} {2021})},\ \Eprint {http://arxiv.org/abs/2107.08451}
  {arXiv:2107.08451 [hep-ph]} \BibitemShut {NoStop}%
\bibitem [{\citenamefont {Ortega}\ \emph {et~al.}(2018)\citenamefont {Ortega},
  \citenamefont {Segovia}, \citenamefont {Entem},\ and\ \citenamefont
  {Fern\'andez}}]{Ortega:2017qmg}%
  \BibitemOpen
  \bibfield  {author} {\bibinfo {author} {\bibfnamefont {P.~G.}\ \bibnamefont
  {Ortega}}, \bibinfo {author} {\bibfnamefont {J.}~\bibnamefont {Segovia}},
  \bibinfo {author} {\bibfnamefont {D.~R.}\ \bibnamefont {Entem}}, \ and\
  \bibinfo {author} {\bibfnamefont {F.}~\bibnamefont {Fern\'andez}},\ }\href
  {\doibase 10.1016/j.physletb.2018.01.005} {\bibfield  {journal} {\bibinfo
  {journal} {Phys. Lett. B}\ }\textbf {\bibinfo {volume} {778}},\ \bibinfo
  {pages} {1} (\bibinfo {year} {2018})},\ \Eprint
  {http://arxiv.org/abs/1706.02639} {arXiv:1706.02639 [hep-ph]} \BibitemShut
  {NoStop}%
\bibitem [{\citenamefont {Entem}\ \emph {et~al.}(2016)\citenamefont {Entem},
  \citenamefont {Ortega},\ and\ \citenamefont
  {Fernández}}]{doi:10.1063/1.4949442}%
  \BibitemOpen
  \bibfield  {author} {\bibinfo {author} {\bibfnamefont {D.~R.}\ \bibnamefont
  {Entem}}, \bibinfo {author} {\bibfnamefont {P.~G.}\ \bibnamefont {Ortega}}, \
  and\ \bibinfo {author} {\bibfnamefont {F.}~\bibnamefont {Fernández}},\
  }\href {\doibase 10.1063/1.4949442} {\bibfield  {journal} {\bibinfo
  {journal} {AIP Conference Proceedings}\ }\textbf {\bibinfo {volume} {1735}},\
  \bibinfo {pages} {060006} (\bibinfo {year} {2016})},\ \Eprint
  {http://arxiv.org/abs/https://aip.scitation.org/doi/pdf/10.1063/1.4949442}
  {https://aip.scitation.org/doi/pdf/10.1063/1.4949442} \BibitemShut {NoStop}%
\bibitem [{\citenamefont {Zyla}\ \emph {et~al.}(2020)\citenamefont {Zyla} \emph
  {et~al.}}]{ParticleDataGroup:2020ssz}%
  \BibitemOpen
  \bibfield  {author} {\bibinfo {author} {\bibfnamefont {P.~A.}\ \bibnamefont
  {Zyla}} \emph {et~al.} (\bibinfo {collaboration} {Particle Data Group}),\
  }\href {\doibase 10.1093/ptep/ptaa104} {\bibfield  {journal} {\bibinfo
  {journal} {PTEP}\ }\textbf {\bibinfo {volume} {2020}},\ \bibinfo {pages}
  {083C01} (\bibinfo {year} {2020})}\BibitemShut {NoStop}%
\bibitem [{\citenamefont {Yan}\ \emph {et~al.}(2021)\citenamefont {Yan},
  \citenamefont {Peng}, \citenamefont {S\'anchez~S\'anchez},\ and\
  \citenamefont {Pavon~Valderrama}}]{Yan:2021tcp}%
  \BibitemOpen
  \bibfield  {author} {\bibinfo {author} {\bibfnamefont {M.-J.}\ \bibnamefont
  {Yan}}, \bibinfo {author} {\bibfnamefont {F.-Z.}\ \bibnamefont {Peng}},
  \bibinfo {author} {\bibfnamefont {M.}~\bibnamefont {S\'anchez~S\'anchez}}, \
  and\ \bibinfo {author} {\bibfnamefont {M.}~\bibnamefont {Pavon~Valderrama}},\
  }\href {\doibase 10.1103/PhysRevD.104.114025} {\bibfield  {journal} {\bibinfo
   {journal} {Phys. Rev. D}\ }\textbf {\bibinfo {volume} {104}},\ \bibinfo
  {pages} {114025} (\bibinfo {year} {2021})},\ \Eprint
  {http://arxiv.org/abs/2102.13058} {arXiv:2102.13058 [hep-ph]} \BibitemShut
  {NoStop}%
\bibitem [{\citenamefont {Gamermann}\ and\ \citenamefont
  {Oset}(2007)}]{Gamermann:2007fi}%
  \BibitemOpen
  \bibfield  {author} {\bibinfo {author} {\bibfnamefont {D.}~\bibnamefont
  {Gamermann}}\ and\ \bibinfo {author} {\bibfnamefont {E.}~\bibnamefont
  {Oset}},\ }\href {\doibase 10.1140/epja/i2007-10435-1} {\bibfield  {journal}
  {\bibinfo  {journal} {Eur. Phys. J. A}\ }\textbf {\bibinfo {volume} {33}},\
  \bibinfo {pages} {119} (\bibinfo {year} {2007})},\ \Eprint
  {http://arxiv.org/abs/0704.2314} {arXiv:0704.2314 [hep-ph]} \BibitemShut
  {NoStop}%
\bibitem [{\citenamefont {Wang}(2021)}]{Wang:2020dgr}%
  \BibitemOpen
  \bibfield  {author} {\bibinfo {author} {\bibfnamefont {Z.-G.}\ \bibnamefont
  {Wang}},\ }\href {\doibase 10.1142/S0217751X21501074} {\bibfield  {journal}
  {\bibinfo  {journal} {Int. J. Mod. Phys. A}\ }\textbf {\bibinfo {volume}
  {36}},\ \bibinfo {pages} {2150107} (\bibinfo {year} {2021})},\ \Eprint
  {http://arxiv.org/abs/2012.11869} {arXiv:2012.11869 [hep-ph]} \BibitemShut
  {NoStop}%
\end{thebibliography}%

\end{document}